\documentclass[10p,onecolumn]{IEEEtran}
[transmag]
\ifCLASSINFOpdf
   \usepackage[pdftex]{graphicx}
\else
\fi
\usepackage{amsmath}

\begin{document}
%
\title{Darwin Among the Cryptocurrencies}


\author{\IEEEauthorblockN{Bernhard K. Meister\IEEEauthorrefmark{1}, and
Henry C.W. Price\IEEEauthorrefmark{2}}
\\
\IEEEauthorblockA{\IEEEauthorrefmark{1}MK2 Finance, Vienna, Austria}
\\
\IEEEauthorblockA{\IEEEauthorrefmark{2}Department of Physics \&  Centre for Complexity Science, Imperial College London, London, SW7 2AZ, UK}
\thanks{
Corresponding author: B.K.~Meister  \,(email: bernhard.k.meister@gmail.com)}}



%



\IEEEtitleabstractindextext{%
\begin{abstract}
 The paper highlights some commonalities between the development of cryptocurrencies and the evolution of ecosystems. 
  Concepts from evolutionary finance embedded in toy models consistent with stylized facts are employed to understand what survival of the fittest means in cryptofinance.
 Stylized facts for ownership, trading volume and market capitalization of cryptocurrencies  are 
 selectively presented in terms of scaling laws. 
\end{abstract}

}

\maketitle

\IEEEdisplaynontitleabstractindextext

%
\IEEEpeerreviewmaketitle

\section{Introduction}
%
%
%
%

 

\IEEEPARstart{C}{ryptocurrencies} with Bitcoin 
in the lead
have grown dramatically over the last decade. Due to isolation from
legacy finance, as described in ``Yields: The Galapagos Syndrome Of Cryptofinance''\cite{galapagos}, the 
colourful world of cryptofinance has evolved  new-fangled products, e.g. stable coins, popularised others, e.g. perpetual futures, and created the field of decentralised finance, where financial transactions are handled largely autonomously\footnote{The involvement of oracles or similar mechanisms acting as as anchor can provide a link to the external world.} by smart contracts.


Darwin  in ``On the Origin of Species''\cite{darwin} established evolutionary biology based on natural selection. 
Competition\footnote{Competition can be on the level of genes, individual organisms or groups.   `Nature, red in tooth and claw' can be supplemented or even replaced by cooperation as was suggested by Kropotkin and  followers.}  for resources  creates winners and losers. Winners have an `edge' and populate the world. As the environment evolves, partly through feedback from organisms themselves, conditions change and continuous adaption  becomes necessary for survival. Occasionally, as in the Cambrian explosion, the process speeds up. 

Over the last decade the world has witnessed an acceleration in the development of money, as a means of exchange, store of value and unit of account.  
Unlike mutations in the natural world, which are random and unguided, changes in protocols of cryptocurrencies are directed and designed by often rational actors to improve fitness.
A multitude of projects exist worldwide 
to find technologically adapted answers to the old question: What is money?

Gregory Chaitin has argued that `nature is programming without a programmer'\cite{chaitin}, and in analogy one could claim that 
`cryptofinance is financial protocol development without central coordination'. 

Cryptocurrencies compete for investor flows, developer commitments  and, where possible, for decentralised finance (DeFi) applications. 
The winners have larger market capitalisation, daily turnover, and are listed on more centralised and decentralised exchanges (CEXs and DEXs). They are also more prominent on social media and often attract 
a larger community of developers. 

In section two evidence is presented for the existence of scaling  laws  for cryptocurrencies. In section three the price impact difference between centralised and decentralised exchanges is  described.
In section four toy models are employed to derive stylised facts with implications for the competitiveness of cryptocurrency protocols. 
In the conclusion the similarity between cryptofinance and biology is again remarked upon. 

\section{Scaling Relationships for Cryptocurrencies}

\IEEEPARstart{I}n this section scaling laws for cryptocurrencies will be studied. 
Scaling laws establish relationships between quantities over several orders of magnitude. In biology for example, since the pioneering work of Kleiber\cite{kleiber, kleiber1} in the early 1930s,  allometric scaling laws, e.g. the $3/4$ power law relating body mass to metabolic rate, have been observed and  multiple derivations\cite{west} have been proposed. It is almost always contentious to verify the existence of a scaling law, since
 data often lacks granularity, changes with time and exponents, due to their sensitivity to the tails of the distribution, are notoriously hard to estimate.
 Therefore,  in this section just  some  suggestive  evidence for the presence of scaling laws in cryptofinance is given. 
 
\noindent
\begin{figure}[h]
\includegraphics[width=\columnwidth]
{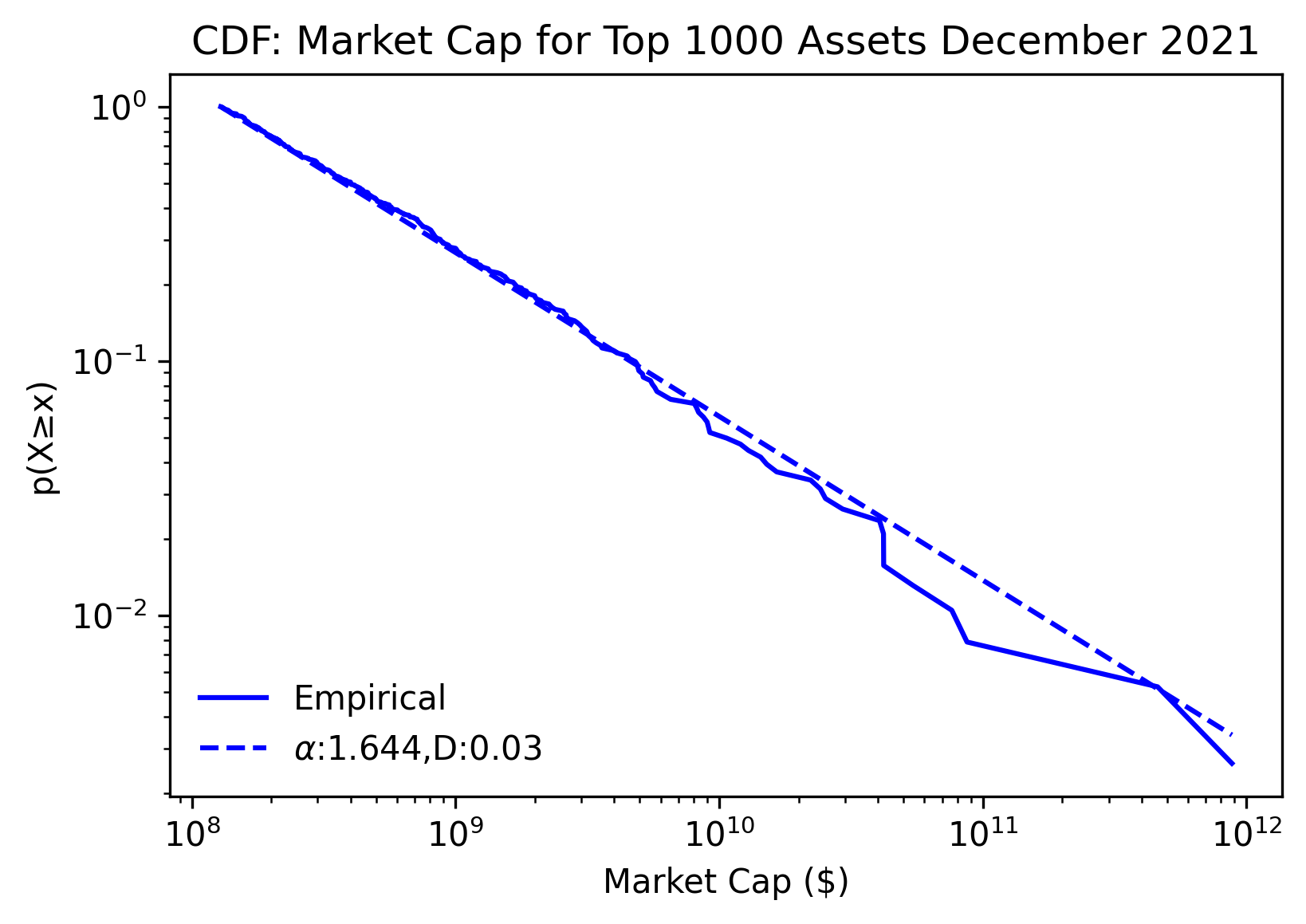}
\caption{Scaling law for the market capitalization of cryptocurrencies. Data to be found on https://coinmarketcap.com. API for current data is used. }
\end{figure}\noindent
 The first figure shows how market capitalisation scales with rank. The biggest cryptocurrency by market capitalization is Bitcoin with  around $\$ 1$ trillion, which is obtained by multiplying coins issued with the current traded price in USD. 
 The power law for the biggest one thousand cryptocurrencies ranked by size, i.e.
 \begin{eqnarray}
C_n \sim K n^{-\alpha} \nonumber
\end{eqnarray}
 with $C_n$ the market capitalisation of the $n$-th cryptocurrency, $K$ the market capitalisation of Bitcoin, and $\alpha$ the scaling law exponent, gives for current market data  (as of 19th December 2021) an exponent of $1.644$. The Python package\cite{Alstott} was employed.


\noindent
\begin{figure}[h]
\includegraphics[width=\columnwidth]
{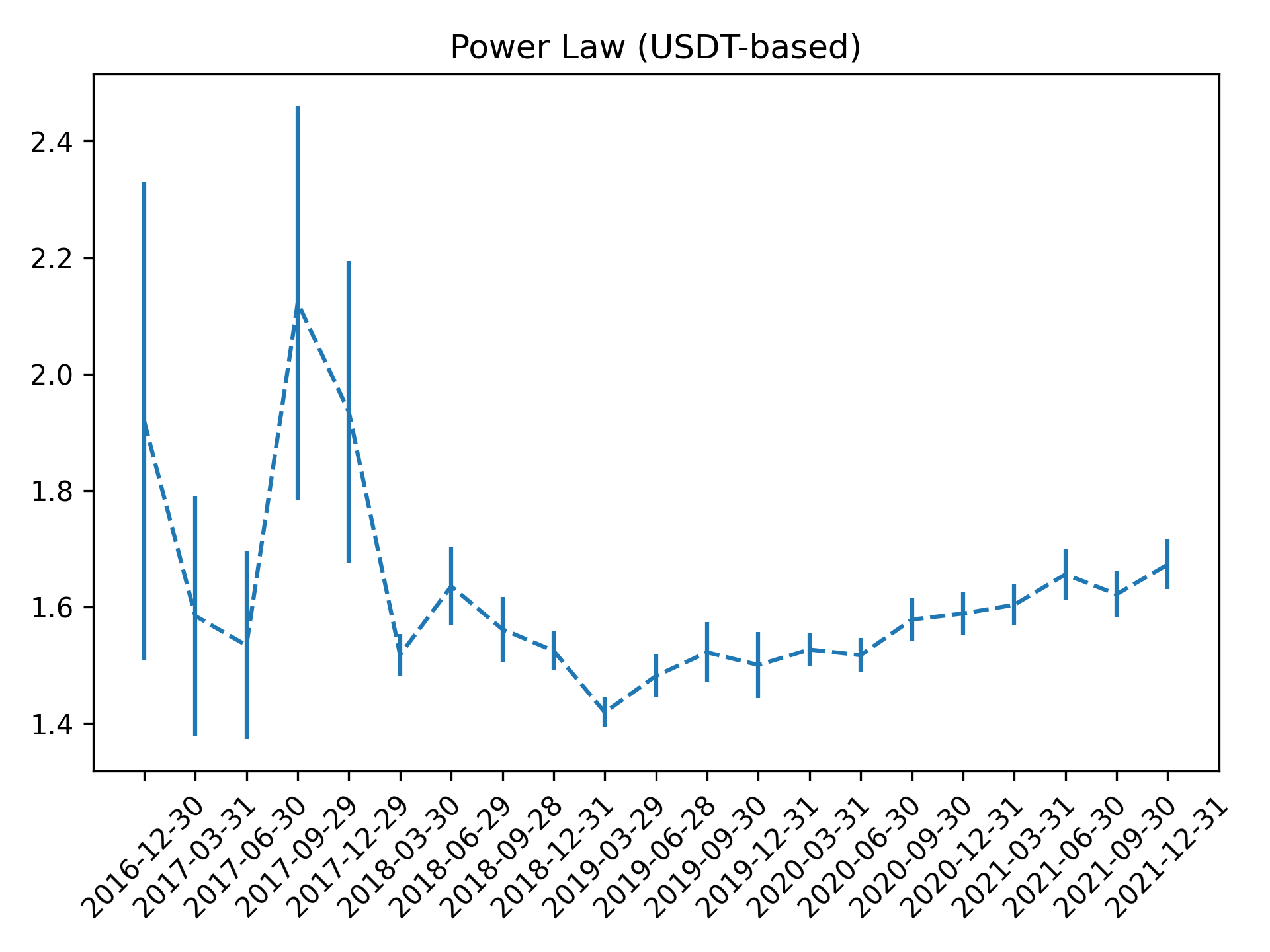}
\caption{ Scaling law for Traded Volume (USD) over 3 month periods, one $\sigma$ error bars,  with USDT as the base currency. Source:{https://www.coinapi.io}}
\end{figure}\noindent
In figure 2 shows a scaling law for the traded volume of cryptocurrencies.
The currencies are binned by quarterly traded volume against Tether (USDT) as a base currencies going back to September 2016. A scaling law is again fitted using the same Python package\cite{Alstott}.
Similar results are obtained for the trading pairs with USD, USD Coin (USDC) and EUR. The number of assets pairs involving USDT has rapidly increased from 11 in September 2016 to the current number of almost 4000. Other currencies have seen a less rapid increase in the number of trading pairs. 
This increase of data has also led to an improvement in the estimation accuracy of the scaling exponent, which is currently around $1.66$. 

Similar scaling laws hold true for NFTs\cite{nft} as well, e.g. scaling of
number of sales per NFT or number of assets per collection seem to scale both close to $3/2$  ($1.4$ and $1.5$).

The paper by Dreyer\cite{dreyer} has a concise derivation of some ratios often found in scaling. If the dimension is $D$, then the common scaling ratio is $D/(D+1)$. As an example, in two dimension the exponent $2/3$ describes how the flow of a river scales with the size of its river basin fed  evenly by rainfall. 

This suggests that one can back out the dimension of a market from the scaling exponent.
If one takes the value of $1.64$, then the associated dimension is close to the non-integer value $3/2$. 
What does this mean?
Cryptocurrencies range widely over protocol space. As the number of cryptocurrencies increase, competition for the most favourite design amplifies. Environmental changes require adaptation. Changes can be soft or hard forks or complete protocol reformulations. One particular prominent alteration is Ethereum moving from proof-of-work to proof-of-stake. 
Investor capital flows into the most appealing cryptocurrencies. Backers  of cryptocurrencies have therefore an incentive to create a favourable environment. This will be discussed in more detail in the fourth section.

Bounds on the scaling exist. If for example the market capitalisation decreases slower than $n^{-1}$, then the combined market capitalisation will theoretically diverge. 

\noindent
\begin{figure}[h]
\includegraphics[width=\columnwidth]
{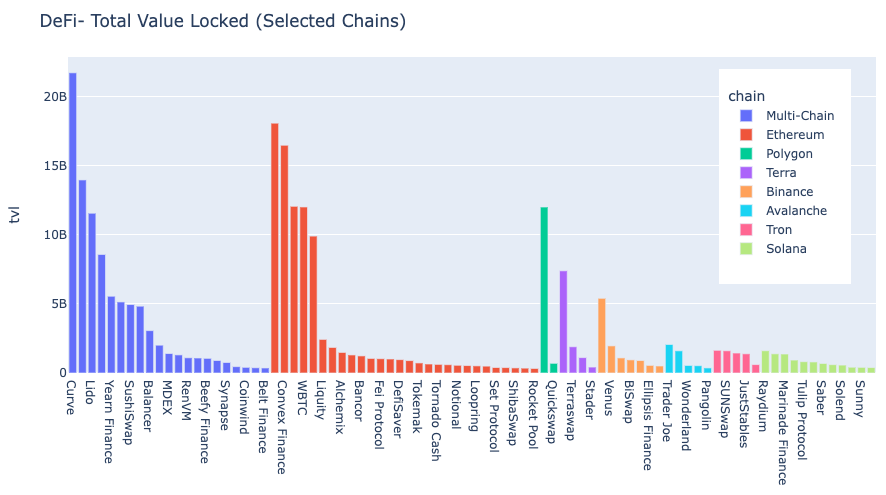}
\caption{ Total Value Locked (TVL ) for different chains and different providers;
Source:{https://www.DefiLlama.com} }
\end{figure}\noindent
Scaling behaviour also exists in  various areas of decentralised finance. An example is Figure 3, which shows Total Value Locked (TVL) for  various Chains. Each colour represents one Chain.
Scaling behaviour is clearly visible. 

In the next section we develop a component useful for toy models in cryptofinance.
\section{Price Impact: DEX vs CEX}
\IEEEPARstart{I}{n} this section the price impact on centralised and decentralised exchanges is studied. It will be shown that CEXs have square root price impact, while for DEXs it is  linear.

 The impact function consistent with dimensional analysis\cite{squareroot, squareroot1}  and backed up by numerous empirical studies of CEXs across a wide range of asset classes has the form
\begin{eqnarray}
 \Delta y  \sim \sigma \sqrt{\Delta x/V}, \nonumber
\end{eqnarray}
  with $\Delta y$ the price change,  $\Delta x$ the impact producing combination of orders, $V$
  the average daily total volume, and  $\sigma$  the `current' daily asset volatility.  
  The average daily total volume is likely to be dominated by non-directional residual noise trading.
  If noise trading is Gaussian, then the imbalance between daily buy and sell orders will be of the order of the square-root of the total noise trading volume and contributes to $\Delta x$.
  One can subdivide $\Delta x$ into  external money flowing into  crypto  and  internal money, which is redeployed due to portfolio updating  linked to for example Kelly optimization as discussed in the next section, plus the spill-over from  noise trading. 

The formula above contrasts with the price impact associated with a DEX based on an Automatic Market Maker (AMM) with a Constant Function Market Maker (CFMM). The most popular form of currency exchange AMM is encapsulated in the formula $K=xy$, where  $x$ and $y$ are the tokens deposited in the pool and $K$ is the resulting constant. Ignoring transaction cost, a switch of one token for another updates $x$ and $y$ in the pool, while leaving the product $K$ unchanged. For details see the Whitepaper\cite{CFMM} or the burgeoning literature.
A conversion of $\Delta x$ changes $y$ to $y'$, and can written as
\begin{eqnarray}
K=(x + \Delta x )(y')\nonumber
\end{eqnarray}
with
\begin{eqnarray}
y'=K/(x + \Delta x )\nonumber
\end{eqnarray}
and
\begin{eqnarray}
y-y'&=&K/x-K/(x + \Delta x )\nonumber\\
&=&K/x(1-  (1-\Delta x/x+(\Delta x/x)^2)+O\big((\Delta x)^3\big)\nonumber\\
&=&K/x(\Delta x/x+(\Delta x/x)^2)+O\big((\Delta x)^3\big)\nonumber
\end{eqnarray}
leading to a price impact, if one sets $\Delta y=y-y'$,  of
\begin{eqnarray}
\Delta y&=&K \Delta x/x^2+ O\big((\Delta x)^2\big),\nonumber
\end{eqnarray}
which is linear in lowest order.

The  difference in market impact, i.e.  between   square root and linear impact in the first order approximation in the CEX and DEX markets respectively, is observable. It can be  used to determine which type of market dominates in a particular currency pair, and  in a triangle currency relationship with a combination of impact functions it can be exploited  for arbitrage. 
This will be further explored in a separate paper, e.g.  Parrondo's paradox\footnote{For more information one can consult the relevant Wikipedia article.}, where the mixing of two `fair' processes, i.e. each behaving like a martingale, can produce, if combined judiciously, a biased process. This effect could be achieved by creating price processes in different currencies pairs through the buying and selling of tokens using CEXs and DEXs. 

In the next section   a series of expanding toy models consistent with the price impact described are introduced.

\section{Stylised facts and Toy models}
\IEEEPARstart{E}{P. Wigner} 
 asked about the `unreasonable effectiveness of mathematics in physics', whereas in other sciences almost the opposite seems to be true, and according to I.M. Gelfand, what is even more surprising is  the unreasonable ineffectiveness of mathematics in biology. Phenomenological descriptions of biological systems exist but few reliable bottom-up theoretical models. This is maybe why  biology is sometimes considered an emergent and not an axiomatic science\footnote{The role of `emergence' has been considered in physics. P.W. Anderson spoke about `more is different' and R.B. Laughlin has emphasized the necessity to consider fundamental laws of physics to be emergent.}.
Is finance, and in particular cryptofinance,  closer to physics or biology?

Cryptofinance with seemingly unending experimentation has similarities to biology, but in this section it will be shown  that toy models also exist consistent with stylised facts, which include the various observed scaling relationships of the   second section and the price impact relationships of the last section. This suggests one cannot only observe scaling laws, but also to a limited extent derive them from first principle. 

An incremental approach is followed, as we slowly refine the toy model 
to progressively reproduce 
the observed stylised facts. This work is speculative and will be extended in subsequent publications.

In each version of the toy models investors are confronted with an optimal allocation problem. How should they, after having decided to invest a certain amount into the crypto asset class, allocate their funds between cryptocurrencies?
Other more advanced strategies associated with DeFis including lending or staking are not considered. 

In the simplest toy model  the market is deemed opaque and investors only have access to historical price data. If this is the case, then investors, who want to optimise their portfolio growth, can  employ the Kelly criterion\cite{kelly1956new}. A wealth of information  about the Kelly criterion can be found in \cite{ziemba}.  If the prices of the cryptocurrencies versus the investors numeraire are normally distributed, then drift, volatility and the co-variance matrix suffices to obtain the optimal fraction. For a single asset  the formula  is
\begin{eqnarray}
\frac{\mu-r}{\sigma^2},\nonumber
\end{eqnarray}
 with $\mu$ the estimated drift, $r$ the riskless interest rate and $\sigma$ the volatility. The extension to multiple assets can be found in \cite{lv1}.
 This suggests an advantage for more established cryptocurrencies, since especially the estimate of the drift improves slowly with the duration of  observation\footnote{Disappointingly, to have sufficient confidence in the drift centuries of daily observation are required.}.  Due to an asymmetry in risk it is sensible to err on the side of caution when choosing the optimal fraction. For more details see the sensitivity analysis of Lv {\it et al.}\cite{lv2}.
The framework can be extended to general Levy distributions of particular relevance to volatile assets like cryptocurrencies, see \cite{bkm2016}.

As an aside, the reliance on past performance can be used to justify the underpricing of IPOs and ICOs, since the initial `pop' in the price is incorporated into the drift estimation and thereby benefits asset allocation. 

As a next step we consider in a qualitative way, how ownership concentration might affect the price trajectory. 
This approach is rooted in the work of Knuteson\cite{knutseon}, who in a series of papers stretching over five years forcefully argued that stock markets around the world are possibly subject to manipulation due to ownership concentration paired with unequal price impact during different periods of the day, which leads to distinct return patterns inter- (overnight) versus intra-day. Curiously, the Chinese market due to potentially greater retail  and lower quantitative fund influence does not follow the same pattern.

The suggestive strategy 
can be divided into two parts. The first part is for investors to have large positions in particular stocks, which are held for an extended period.
The second part is for the same investors to add on top a cyclical strategy of buying and selling additional amounts of the same stocks at specific times. Stocks are bought, when the price impact is high, e.g. at the market open, and equal amounts are sold again when the price impact is low, e.g. close to the end of the trading day on major exchanges. This results in an overall increase in the price of the stocks.
The cyclical trading is associated with transaction and impact cost and leads in isolation to a loss. If this is more than off-set by the mark-to-market gains associated with the long position described in part one, then this seems like a (perpetual) money making machine. 

We are agnostic about Knuteson's analysis, since the influence of `unknown unknowns'\footnote{
An Occam's razor based argument could reduce long-running distortions of financial markets exemplified by unusual equity returns to excessive liquidity provided by central banks. 

The role of greed and fear as a hurdle to align the interests of investors is substantial. If a strategy is widely known and relatively obvious, then there will be meta-strategies involving front-running, which leads to hard to predict feedback mechanisms, e.g.  price impact could asymptotically equalize throughout the day.  

}, to use a term popularised by former defense secretary Rumsfeld, prevents easy answers, but  the elegance of the argument is appreciated.

One opens a Pandora's box even without round-robin trading, if one pursues this line of reasoning. 
Does a company for example have a fiduciary duty when repurchasing stock to minimize price impact and as a consequence purchasing cost, or is the aim instead to maximise price impact to raise the value of the company? 

Climbing the market capitalisation rank in crypto is a matter of survival and literally billions await the winners. 
Opportunities for boosterism are amplified in cryptofinance, 
as cryptocurrencies often have  known progenitors,  concentrated ownership and uneven liquidity. 
There is anecdotal evidence that on weekends, when  the legacy financial market is mostly dormant, cryptocurrencies have occasional price gaps. 
 
The effects described above can be quantified, if one estimates price impact, drift,  interest rates and the inverse of the covariance matrix\footnote{The inverse matrix can be replaced by the generalized  Moore-Penrose Inverse.},   and then uses the Kelly criterion to determine what this implies for the optimal investment fractions\cite{bkm2016}. This will be done in a subsequent publication.

A more intricate strategy exploiting differential impact in legacy finance has been ascribed to Citibank. The bank in July 2004 almost simultaneous accumulated  off-setting  position in the European fixed income market. The differential in price-impact was profitably exploited, but the fall-out\footnote{Inside the bank the strategy according to articles in the Wall Street Journal and the Financial Times was given the moniker `Dr. Evil'. When this become public, regulators and the media had a field day.
``Citigroup subsequently paid \$25 million to settle allegations by the U.K. Financial Services Authority (FSA) that the bank failed to supervise its traders and also failed to conduct its business with `due skill, care and diligence'"\cite{wilmarth, wsj}. For further details see the various articles by A.E. Wilmarth Jr\,\,.  The `forced' arbitrage trade was complicated and was spread over multiple exchanges, products and times. A major component was the off-setting of a price shifting multi-billion short position in Eurobonds acquired on the electronic MTS market with a position in the more liquid and lower price impact Eurex bond futures. 
} was unpleasant.

Based on  two relatively weak assumptions, ownership concentration and unequal price impact,  one can imagine scenarios where a cryptocurrency like Bitcoin can turn into a `doomsday machine'. The term is borrowed form the movie `Dr Strangelove' and the book by Daniel Ellsberg\cite{ellsberg}. In the movie an air force general, who has gone rogue, launches an unauthorised limited nuclear attack on the Soviet Union. This triggers the `doomsday machine' to launch automatically an all-out nuclear response.  
As Dr Strangelove explains in the movie: `Of course, the whole point of a `doomsday machine' is lost, if you keep it a  secret!'
For the same reason it is helpful for the market to be aware of potential destabilizing run-a-away strategies\footnote{In the 80ties the nuclear confrontation between the superpowers became increasingly unstable due to technological changes (multi-warhead missiles and SDI) and it was providential that the cold war ended before it got completely out of hand.}. 
 
The manipulation strategy described above could lead to a sequence of instabilities. The full impact is maybe another decade away, since currently Bitcoin has a market capitalisation of less than \$1  trillion, whereas world Gross Domestic Product (GDP) is around \$80 trillion. For  world wealth different amounts are quoted in the literature.  A McKinsey \& Company Global Institute study published in November 2021\cite{mck} gives a figure of $\$ 510$ trillion.
Since Bitcoin is only about  ~1\% of world GDP  and 0.2\% of world wealth there is  still some space for growth.
From the last peak of $\$20k$ per Bitcoin at the end of 2017 to the most recent peak  of $\$ 60k$ per Bitcoin it took about three years.  As a Gedanken-experiment, for Bitcoin to reach a double digit percentage of world wealth it would have to increase by a factor of $50+$, which at a rate  of tripling every three years, 
 would take around a decade. There is therefore maybe ample time
 to watch and possibly intervene into what could be a slow motion `doomsday machine'.
In \cite{bkm2016} it was shown that even without differential price impact price instabilities can occur.

          
%

Up to now we assumed that cryptocurrencies exist in a vacuum by ignoring for example low bandwidth and low signal to noise ratio communication channels between promoters  and investors provided by  Twitter, Discord, Signal and similar services. If utility derived from Github commits, updating speed, transaction cost, miner hash powers and other features is included,  then the optimal investment strategy becomes exceedingly hard to  quantify. 

\section{Conclusion}

\IEEEPARstart{I}{N} this paper some similarities between cryptocurrencies and biological systems were highlighted. 
Scaling laws were described, and toy models constructed. 

Many open questions remain. How are profit and losses distributed between different investors classes? As a comparison, in a recent paper\cite{bike} the shifting ownership of stocks in British bicycle companies during the mania of 1895 to 1900 was analysed. The bicycle company share index first approximately tripled and then fell below the starting value. Investors entered and exited at different times. Their transactions and shareholdings  were recorded in voluminous company ledgers, which in some cases are still available. Informed investors were able to reduce their ownership presciently, while gentleman investors accrued losses. Trading volumes peaked when prices reached their zenith.

The analogy between Darwin's theory and evolutionary cryptofinance can be elaborated. A winning cryptocurrency attracts capital due to a superior Sharpe ratio and other attractive features of the protocol. This is 
not a static but a dynamic problem, as promoters try to obtain an advantage by guiding the asset price and by enhancing  the protocol.

A speculative comment about possible developments in cryptocurrencies rounds of the paper.  It could be called the `Red Queen interpretation’ of cryptocurrencies, since the Red Queen in Lewis Carroll’s ``Through the Looking-Glass" commands Alice to perform “all the running you can do, to keep in the same place. If you want to get somewhere else, you must run at least twice as fast”. To keep abreast with competitors cryptocurrency protocols have continuously to adapt and any stumble might prove fatal.

\section*{Acknowledgment}
\noindent
We gratefully acknowledge conversations with DC Brody, TS Evans and R Kelly. 
The title of the paper was inspired by ``Darwin among the machines'', a newspaper article of 1863 by Samuel Butler, and a book of 1998 by George Dyson.

\ifCLASSOPTIONcaptionsoff
  \newpage
\fi



%

 \end{document}